\documentstyle[amssymb,aps,12pt]{revtex}
\begin{document}
\title{Gauged Dimension Bubbles}
\author{E.I. Guendelman\thanks{%
E-mail: guendel@bgumail.ac.il}}
\address{{\it Department of Physics, Ben-Gurion University of the Negev,}\\
{\it Box 653, IL-84105 Beer Sheva, Israel}}
\author{J.R. Morris\thanks{%
E-mail: jmorris@iun.edu}}
\address{{\it Physics Dept., Indiana University Northwest,}\\
{\it 3400 Broadway, Gary, Indiana 46408 USA}\\
\smallskip\ }
\maketitle

\begin{abstract}
Some of the peculiar electrodynamical effects associated with gauged
``dimension bubbles'' are presented. Such bubbles, which effectively enclose
a region of 5d spacetime, can arise from a 5d theory with a compact extra
dimension. Bubbles with thin domain walls can be stabilized against total
collapse by the entrapment of light charged scalar bosons inside the bubble,
extending the idea of a neutral dimension bubble to accomodate the case of a
gauged U(1) symmetry. Using a dielectric approach to the 4d dilaton-Maxwell
theory, it is seen that the bubble wall is almost totally opaque to photons,
leading to a new stabilization mechanism due to trapped photons. Photon
dominated bubbles very slowly shrink, resulting in a temperature increase
inside the bubble. At some critical temperature, however, these bubbles
explode, with a release of radiation.

\end{abstract}

%\date{}

\section{Introduction}

An inhomogeneous higher dimensional spacetime compactified to four
dimensions (4d) can contain pockets, or, what may be referred to as
``dimension bubbles'', where the extra dimensional scale factor becomes
large enough that the spacetime has an effective dimensionality within the
pocket that is higher than in the surrounding four dimensional spacetime
outside of it\cite{DG,BG}. Here we consider a dimension bubble, arising from
the dimensional reduction of a 5d theory, which encloses a 5d region and is
surrounded by a region that is effectively 4d, i.e., the extra dimensional
scale factor changes rapidly from the interior of the bubble to the
exterior. The two regions are, from a 4d perspective, separated by a domain
wall generated by a scalar field associated with the scale factor of the
extra space dimension. Such a domain wall can result from a Rubin-Roth
potential\cite{RR} (which includes bosonic as well as fermionic Casimir
energies--which can stabilize the compact extra dimension from collapsing
due to the gravitational Casimir effect\cite{AC}) along with a cosmological
constant\cite{BG}, or, in the case of more than one extra dimension, from
higher dimensional Maxwell fields\cite{CGHW} or their generalizations\cite
{DG}. For simplicity, we take the 4d spacetime to be Minkowski and the extra
space dimension to be toroidally compact, so that the 5d spacetime has the
topology of $M_4\times S^1$. (In this sense, the dimension bubbles studied
here are simplifications of ``gravitational bags'', previously analyzed by
Davidson and Guendelman\cite{DG}, which have an associated nontrivial
spacetime geometry.) Although gravitational bags are static solutions where
the surface tension pushing the bubble wall inward is equilibrated with a
nontrivial pressure from the scalar field inside, the scalar field inside
suffers from a singular behavior at the center of the geometry (although
there is still finite energy due to a gravitational effect). There is,
however, another way to stabilize the bubble which does not require a
singular scalar field at the center of the geometry. The stabilization can
be achieved if the inside of the bubble is filled with scalar bosons
described by a complex scalar field $\chi $\cite{JM}.

Previously, it was assumed that the $\chi $ field only possessed a global $%
U(1)$ symmetry\cite{JM} giving rise to a conserved number charge $Q$. Here,
we consider the case where the scalar field $\chi $ has a {\it local} $U(1)$
symmetry, i.e., a $U(1)$ ``gauged dimension bubble''. Because the extra
dimensional scale factor $B(x)$ can take drastically different values inside
and outside of the bubble, the effective ``dielectric function'' of these
two regions can also have drastically different values. As a result, there
are nontrivial electromagnetic effects associated with the bubble wall and
its 5d interior. In particular, the EM contribution to the mass $M$ of the
bubble is reduced from what would be expected if the bubble interior were
also 4d, that is, if the dilaton associated with the scale factor $B$ were
constant everywhere. In addition, the bubble wall is found to be almost
perfectly reflecting to photons, so that photons cannot pass through the
wall from either direction. Entrapped photons can themselves stabilize a
bubble from total collapse. The effects of the extra dimension therefore
make the gauged dimension bubble an object that is quite different from
other purely 4d nontopological solitons\cite{NTS}, such as gauged Q balls%
\cite{Q1,Q2,Q3}, charged vacuum bubbles\cite{JM1}, and Fermi balls\cite
{MC,FB1,FB2,FB3}, that have been studied previously.

We first present the dielectric approach to the study of electromagnetic
effects of dimension bubbles, considering a bubble with a 5d interior
(emerging from a 5d theory dimensionally reduced to 4d) as a specific
prototype. This dielectric approach is then applied to the case of a charged
bubble having a conserved number $Q$ of charged $\chi $ bosons trapped
inside the bubble. Upon evaluating the various contributions to the bubble's
energy, the bubble's equilibrium radius $R$ and mass $M$ can be obtained. We
shall also make a couple of simplifying assumptions. First, we assume a thin
walled bubble, i.e., the wall thickness $\delta \ll R$, so that in a
simplifying limit we may take the inner radius $R_{-}$ and outer radius $%
R_{+}$ of the wall to coincide, $R_{-}$, $R_{+}\rightarrow R$. It is within
the wall that the scale factor $B(x)$ varies rapidly, and we assume that $B$
takes on different constant values in the interior and exterior regions,
with the interior value being much greater than the exterior one, $B_{in}\gg
B_{out}$. This allows for an efficient mechanism of trapping the $\chi $
bosons inside the bubble, since the boson mass ($m_\chi \propto B^{-1/2}$),
assumed to be small inside, becomes very large outside, $m_{\chi ,in}\ll
m_{\chi ,out}$, and a boson therefore experiences an enormous force $\vec{F}%
\sim -\nabla m_\chi $ exerted on it by the wall, keeping it in the interior
of the bubble.

\section{The Model}

\subsection{Metric Ansatz and Dimensional Reduction}

We begin with a 5d action of the general form 
\begin{equation}
S_5=%
%TCIMACRO{\dfrac 1{2\kappa _{N(5)}^2} }
%BeginExpansion
{\displaystyle {1 \over 2\kappa _{N(5)}^2}}
%EndExpansion
\int d^5x\sqrt{\tilde{g}_5}\left\{ \tilde{R}_5-2\Lambda +2\kappa _{N(5)}^2%
{\cal L}_5\right\}  \label{eq1}
\end{equation}

\noindent that is defined on a 5d spacetime with a metric described by 
\begin{equation}
ds^2=\tilde{g}_{MN}dx^Mdx^N=\tilde{g}_{\mu \nu }dx^\mu dx^\nu +\tilde{g}%
_{55}dy^2  \label{eq2}
\end{equation}

\noindent Here, $x^M=(x^\mu ,y)$ is the set of coordinates on a 5d spacetime
with topology of $M_4\times S^1$ that has a toroidally compact spacelike
extra dimension $x^5=y$, which is assumed to be a linear coordinate lying
within the range $0\leq y\leq 2\pi R$. We use the indices $M$, $N=0,\cdot
\cdot \cdot ,3,5$ to label the coordinates of the 5d spacetime and the
indices $\mu $, $\nu =0,\cdot \cdot \cdot ,3$ to label the coordinates of
the noncompact 4d spacetime. A zero mode Kaluza-Klein ansatz is assumed
where the fields and metric are independent of $x^5$, {\it i.e.}, $\tilde{g}%
_{MN}=\tilde{g}_{MN}(x^\mu )$ depends only on the 4d coordinates with $%
\partial _5\tilde{g}_{MN}=0$. We also assume that the metric factorizes so
that $\tilde{g}_{\mu 5}=0$. A dimensionless scale factor $B(x^\mu )$, with $%
\tilde{g}_{55}=-B^2$, is associated with the extra dimension, along with an
associated scalar field $\varphi $ that is related to $B$ by 
\begin{equation}
\varphi =\frac 1{\kappa _N}\sqrt{\frac 32}\ln B  \label{eq3}
\end{equation}

\noindent The constant $\kappa _N$ is related to the 4d Planck mass $M_P$ by 
$\kappa _N=\sqrt{8\pi G}=\sqrt{8\pi }M_P^{-1}$. The extra dimensional scale
factor can then be written as $B=e^{\sqrt{2/3}\,\kappa _N\varphi }$.

Determinants of the 4d and 5d parts of the original metric $\tilde{g}_{MN}$
are denoted by $\tilde{g}=\det \tilde{g}_{\mu \nu }$ and $\tilde{g}_5=\det 
\tilde{g}_{MN}$, respectively, so that $\,\sqrt{|\tilde{g}_5|}=\sqrt{-\tilde{%
g}}\sqrt{|\tilde{g}_{55}|}$. In the action given by (\ref{eq1}), $\kappa
_{N(5)}=\sqrt{8\pi G_5}$ represents a 5d gravitational constant, $\tilde{R}%
_5=\tilde{g}^{MN}\tilde{R}_{MN}$ is the 5d Ricci scalar built from $\tilde{g}%
_{MN}$, ${\cal L}_5$ is a Lagrangian of the 5d theory, and $\Lambda $ is a
5d cosmological constant. In order to pass to an effective 4d theory we
define ${\cal L}=(2\pi R){\cal L}_5$ and define $\kappa _{N(5)}$ in terms of 
$\kappa _N$ by $\kappa _{N(5)}^2=(2\pi R)\kappa _N^2$.

The 4d Jordan Frame metric is $\tilde{g}_{\mu \nu }$, the $\mu \nu $ part of
the 5d metric $\tilde{g}_{MN}$. A 4d Einstein Frame metric $g_{\mu \nu }$
can be defined by $g_{\mu \nu }=B\tilde{g}_{\mu \nu }=e^{\sqrt{2/3}\kappa
_N\varphi }\tilde{g}_{\mu \nu }$ and the line element in (\ref{eq2}) can be
rewritten in terms of the Einstein Frame metric and extra dimensional scale
factor as

\begin{equation}
\begin{array}{ll}
ds^2 & =B^{-1}g_{\mu \nu }dx^\mu dx^\nu -B^2dy^2 \\ 
& =e^{-\sqrt{2/3}\,\kappa _N\varphi }g_{\mu \nu }dx^\mu dx^\nu -e^{2\sqrt{2/3%
}\,\kappa _N\varphi }dy^2
\end{array}
\label{eq4}
\end{equation}

\noindent Using (\ref{eq4}), a dimensional reduction of the action given by (%
\ref{eq1}) gives the effective 4d Einstein Frame action 
\begin{equation}
S=\int d^4x\sqrt{-g}\left\{ \frac 1{2\kappa _N^2}R+\frac 12(\nabla \varphi
)^2+e^{-\sqrt{2/3}\kappa _N\varphi }[{\cal L}-\frac 1{\kappa _N^2}\Lambda
]\right\}  \label{eq5}
\end{equation}

\noindent where $R=g^{\mu \nu }R_{\mu \nu }$ is the 4d Ricci scalar built
from the 4d Einstein Frame metric $g_{\mu \nu }$ and $g=\det g_{\mu \nu }$.

\subsection{4d (Einstein Frame) Effective Lagrangian}

Consider a Lagrangian from the 5d theory of a $U(1)$ gauged scalar field $%
\chi $, 
\begin{equation}
{\cal L}=(2\pi R){\cal L}_5=-\frac 14\ \tilde{F}^{\prime MN}\tilde{F}%
_{MN}^{\prime }+(\tilde{D}^M\chi )^{*}(\tilde{D}_M\chi )-V(|\chi |)
\label{e6}
\end{equation}

\noindent which gives rise to an effective 4d (EF) Lagrangian ${\cal L}%
_4=B^{-1}{\cal L}$, and 
\begin{equation}
\tilde{F}_{MN}^{\prime }=\partial _MA_N^{\prime }-\partial _NA_M^{\prime
},\,\,\,\,\,\,\,\,\,\,D_M\chi =(\nabla _M+ie_0A_M^{\prime })\chi  \label{e7}
\end{equation}

\noindent with $e_0$ and $A_M^{\prime }$ being the charge parameter and
gauge field potential, respectively, appearing in the original 5d
Lagrangian, and tildes remind us that indices are raised and lowered with
the metric $\tilde{g}_{MN}$, so that $(\tilde{D}^M\chi )^{*}(\tilde{D}_M\chi
)=\tilde{g}^{MN}(D_M\chi )^{*}(D_N\chi )$. It is assumed that $\partial
_5\chi =0$ so that $D_5\chi =ie_0A_5^{\prime }\chi $. We then have, with the
assumption that $\partial _5A_\mu ^{\prime }=0$, 
\begin{equation}
-\frac 14\tilde{F}^{\prime MN}\tilde{F}_{MN}^{\prime }=-\frac 14B^2F^{\prime
\mu \nu }F_{\mu \nu }^{\prime }+\frac 12B^{-1}(\partial ^\mu A_5^{\prime
}\partial _\mu A_5^{\prime })  \label{e8}
\end{equation}

\noindent Using $\tilde{g}^{\mu \nu }=Bg^{\mu \nu }$, $\tilde{g}%
^{55}=-B^{-2} $, the scalar field kinetic term is given by 
\begin{equation}
\begin{array}{cc}
(\tilde{D}^M\chi )^{*}(\tilde{D}_M\chi ) & =(\tilde{D}^\mu \chi )^{*}(\tilde{%
D}_\mu \chi )+(\tilde{D}^5\chi )^{*}(\tilde{D}_5\chi ) \\ 
& =B\left| D_\mu \chi \right| ^2-e_0^2B^{-2}A_5^{\prime 2}\left| \chi
\right| ^2
\end{array}
\label{e9}
\end{equation}

The 4d EF effective Lagrangian ${\cal L}_4=B^{-1}{\cal L}$ then becomes 
\begin{equation}
\begin{array}{ll}
{\cal L}_4 & =-\frac 14BF^{\prime \mu \nu }F_{\mu \nu }^{\prime }+\frac 12%
B^{-2}(\partial A_5^{\prime })^2+\left| D_\mu \chi \right| ^2 \\ 
& -e_0^2B^{-3}A_5^{\prime 2}|\chi |^2-B^{-1}V(|\chi |)
\end{array}
\label{e10}
\end{equation}

\noindent The $\varphi $ - dependent potential $U(\varphi )$ (which
contains, e.g., the Rubin-Roth potential, $V_{RR}$, and cosmological
constant, $\Lambda $, terms), along with the $\varphi $ kinetic term $\frac 1%
2(\partial \varphi )^2$ and gravitational term $\frac 1{2\kappa _N^2}R$ can
be added to get a total 4d effective Lagrangian 
\begin{equation}
\begin{array}{ll}
{\cal L}_{eff} & ={\cal L}_4+\frac 1{2\kappa _N^2}R+\frac 12(\partial
\varphi )^2-U \\ 
& =\frac 1{2\kappa _N^2}R+\frac 12(\partial \varphi )^2-\frac 14BF^{\prime
\mu \nu }F_{\mu \nu }^{\prime }+\frac 12B^{-2}(\partial A_5^{\prime
})^2+\left| D_\mu \chi \right| ^2-W
\end{array}
\label{e11}
\end{equation}

\noindent where we define a total effective potential 
\begin{equation}
W=U(\varphi )+B^{-1}V(|\chi |)+e_0^2B^{-3}A_5^{\prime 2}|\chi |^2
\label{e12}
\end{equation}

\subsection{Dielectric Approach}

The 4d Einstein Frame effective Lagrangian ${\cal L}_{eff}$ above contains a
Maxwell term $-\frac 14BF^{\prime \mu \nu }F_{\mu \nu }^{\prime }$ and a
gauge covariant derivative $D_\mu \chi =(\partial _\mu +ie_0A_\mu ^{\prime
})\chi $. Let us define a rescaled gauge field $A_M=B_4^{1/2}A_M^{\prime }$,
where $B_4$ (or $B_5$) denotes the value of $B$ in a region of 4d (or 5d)
spacetime outside (or inside) of a dimension bubble, so that in a region of
4d spacetime the Maxwell term is properly normalized and assumes a canonical
form, $-\frac 14B_4F^{\prime \mu \nu }F_{\mu \nu }^{\prime }=-\frac 14F^{\mu
\nu }F_{\mu \nu }$, with 
\begin{equation}
F_{\mu \nu }=\partial _\mu A_\nu -\partial _\nu A_\mu ,\,\,\,\,\,\,\,A_\mu
=B_4^{1/2}A_\mu ^{\prime },\,\,\,\,\,\,\,A_5=B_4^{1/2}A_5^{\prime }
\label{a1}
\end{equation}

\noindent The gauge covariant derivative operator $D_\mu =(\partial _\mu
+ie_0A_\mu ^{\prime })$ can be written as 
\begin{equation}
D_\mu =(\partial _\mu +ie_0B_4^{-1/2}A_\mu )=(\partial _\mu +ieA_\mu )
\label{a2}
\end{equation}

\noindent where we have defined the 4d effective, or physical, charge 
\begin{equation}
e=B_4^{-1/2}e_0  \label{a3}
\end{equation}

\noindent We also define the ``dielectric function'' 
\begin{equation}
\kappa (x^\mu )=\frac{B(x^\mu )}{B_4}  \label{a3a}
\end{equation}

In terms of physical 4d quantities, the effective 4d Lagrangian in (\ref{e11}%
) now assumes the form 
\begin{equation}
{\cal L}_{eff}=\frac 1{2\kappa _N^2}R+\frac 12(\partial \varphi )^2-\frac 14%
\kappa F_{\mu \nu }F^{\mu \nu }+\frac 12\kappa ^{-2}B_4^{-3}(\partial
A_5)^2+|D_\mu \chi |^2-W  \label{a4}
\end{equation}

\noindent where 
\begin{equation}
W=U(\varphi )+\kappa ^{-1}B_4^{-1}V(|\chi |)+e^2\kappa
^{-3}B_4^{-3}A_5^2|\chi |^2  \label{a5}
\end{equation}

\noindent The function $\kappa $ is seen to play the role of a dielectric
function, except that in this case it arises as a consequence of the
possible position dependence of the extra dimensional scale factor $B$.
Thus, from a 4d perspective, a region of effectively 5d spacetime is viewed
as being endowed with a dielectric property described by $\kappa $. We also
note that in a 4d region $\kappa \rightarrow 1$, and that $\kappa
_5=B_5/B_4>1$, and, in particular, we assume that $\kappa _5\gg 1$. Using
the familiar results of electrostatics, we have that the normal component of
the ``displacement'' field $\vec{D}=\kappa \vec{E}$ and the tangential
components of the electric field $\vec{E}$ are continuous at an interface
between two media with different dielectric constants.

\subsection{Effective Potential$,$ Vacuum$,$ and $\chi $ Boson Mass}

The total effective potential $W$ of the 4d theory is given by eq. (\ref{a5}%
). The vacuum state is obtained by minimizing $W$ with respect to $\varphi $
(or $\kappa $), $\chi $, and $A_5$: 
\begin{equation}
\begin{array}{ll}
%TCIMACRO{\dfrac{\partial W}{\partial \kappa } }
%BeginExpansion
{\displaystyle {\partial W \over \partial \kappa }}
%EndExpansion
& =%
%TCIMACRO{\dfrac{\partial U}{\partial \kappa } }
%BeginExpansion
{\displaystyle {\partial U \over \partial \kappa }}
%EndExpansion
-B_4^{-1}\kappa ^{-2}V-3e^2B_4^{-3}\kappa ^{-4}A_5^2|\chi |^2=0, \\ 
&  \\ 
%TCIMACRO{\dfrac{\partial W}{\partial \chi ^{*}} }
%BeginExpansion
{\displaystyle {\partial W \over \partial \chi ^{*}}}
%EndExpansion
& =B_4^{-1}\kappa ^{-1}%
%TCIMACRO{\dfrac{\partial V}{\partial \chi ^{*}} }
%BeginExpansion
{\displaystyle {\partial V \over \partial \chi ^{*}}}
%EndExpansion
+e^2B_4^{-3}\kappa ^{-3}A_5^2\chi =0, \\ 
&  \\ 
%TCIMACRO{\dfrac{\partial W}{\partial A_5} }
%BeginExpansion
{\displaystyle {\partial W \over \partial A_5}}
%EndExpansion
& =2e^2B_4^{-3}\kappa ^{-3}A_5|\chi |^2=0
\end{array}
\label{e13}
\end{equation}

\noindent These equations are solved by $A_5=0$, $\partial V/\partial \chi
^{*}=0$, and $\partial U/\partial \kappa -B_4^{-1}\kappa ^{-2}V=0$. (Later
we will consider the case for which $V=\mu _0^2\chi ^{*}\chi $, giving $\chi
=0$, $V=0$, $\partial U/\partial \kappa =0$, with $A_5$ undetermined, and we
will assume that $A_5=0$.)

The 5d mass parameter for the $\chi $ field is denoted by $\mu
_0^2=(\partial ^2V/\partial \chi ^{*}\partial \chi )|_{vac}$, so that the
effective $\chi $ boson mass in the 4d effective (EF) theory is given by 
\begin{equation}
m_\chi ^2=\left( \frac{\partial ^2W}{\partial \chi ^{*}\partial \chi }%
\right) _{vac}=\left[ B^{-1}\mu _0^2+e^2B_4^{-3}\kappa ^{-3}A_5^2\right]
_{vac}  \label{e13a}
\end{equation}

\noindent Assuming the vacuum value of $A_5$ to vanish, we have simply 
\begin{equation}
m_\chi =B^{-1/2}\mu _0=\kappa ^{-1/2}\mu _4  \label{e13b}
\end{equation}

\noindent where we define $\mu _4=B_4^{-1/2}\mu _0$. In this effective 4d
dilaton-Maxwell system, the effects of the scale factor (or dilaton $\varphi 
$) become manifest in the 4d mass parameter $m_\chi $. In the 5d region of a
dimension bubble interior, where $\kappa \gg 1$, the mass $\kappa
_5^{-1/2}\mu _4$ becomes very small or negligible in comparison to the mass $%
\mu _4$ in the 4d region outside the bubble. This effective mass dependence
of $m_\chi $ upon $\kappa $ gives rise to the entrapment of $\chi $ bosons
inside the bubble.

\section{Gauged Dimension Bubble}

Consider the case where a domain bubble forms, entrapping a conserved number 
$Q$ of $\chi $ bosons. (For definiteness, we take the electromagnetic charge 
$Qe$ to be positive and the potential $V$ is taken to be given by $V=\mu
_0^2\chi ^{*}\chi $.) The simplified situation is assumed to exist for which
the bosons are nearly massless inside the bubble, so that a gas of
ultra-relativistic particles exists inside, and the bosons are massive
outside, with the boson mass being $m$. We assume that the bubble takes a
spherical shape at equilibrium. The inner surface of the bubble wall lies at
a radius $R_{-}$ and the outer surface is located at $R_{+}$. We will use a
thin wall approximation, in which case $R_{-}\approx R_{+}\approx R$.

The mass $M$ of the bubble gets contributions from the kinetic energy ${\cal %
E}_\chi $ of the $\chi $ bosons, the energy of the domain wall forming the
bubble surface, ${\cal E}_W=4\pi R^2\sigma $, and the electromagnetic (EM)
energy due to the entrapped $\chi $ bosons, ${\cal E}_{em}$. In addition,
there is a contribution from the $\varphi $ - dependent potential $U(\varphi
)$ in the interior of the bubble. This arises from the fact that we are
considering the case\cite{JM} where $U(\varphi )\rightarrow 0$ in the 4d
vacuum region and $U(\varphi )>0$ in the 5d vacuum region of the bubble's
interior, where $\varphi $ assumes a large, but finite, value. Denoting the
value of $U(\varphi )$ in the bubble's interior by $\lambda $, a constant in
our approximation, we have the corresponding volume term ${\cal E}_V=\frac 43%
\pi \lambda R^3$ contributing to the mass $M$ of the bubble. The bubble mass
can therefore be written as

\begin{equation}
M={\cal E}_\chi +{\cal E}_{em}+{\cal E}_W+{\cal E}_V  \label{e22}
\end{equation}

The first term ${\cal E}_\chi $ in this expression for the mass can be
estimated easily. For the ground state kinetic energy of an
ultra-relativistic $\chi $ boson trapped inside a bubble of radius $R$, we
take the boson wavelength to be roughly equal to the bubble diameter, $%
\lambda _\chi \approx 2R$. Then the kinetic energy is ${\cal E}_{\text{kin}%
}\approx 2\pi /\lambda _\chi \approx \pi /R$. The kinetic energy of $Q$
bosons in the ground state is then approximately ${\cal E}_\chi =Q\pi /R$.
For the EM energy we need to integrate the electromagnetic energy density $u=%
\frac 12\vec{D}\cdot \vec{E}=\frac 12\kappa E^2$ over all space, 
\begin{equation}
{\cal E}_{em}=\int udV=2\pi \int_0^\infty \kappa E^2\,r^2dr  \label{e22a}
\end{equation}

\subsection{Electric and Displacement Fields}

The EM field satisfies the field equation $\nabla _\mu (\kappa F^{\mu \nu
})=j^\nu =eJ^\nu $, where $J^\mu $ is the current per unit charge that
generates the $\chi $ boson number density, so that $Q=\int J^0dV$. As a
simplifying approximation we assume that $B$ and $\kappa =B/B_4$ take on
constant values inside and outside the bubble: 
\begin{equation}
B=\left\{ 
\begin{array}{ll}
B_5, & \text{inside, }(r<R) \\ 
B_4, & \text{outside, }(r>R)
\end{array}
\right\} ,\,\,\,\,\,\kappa =\left\{ 
\begin{array}{ll}
\kappa _5=\frac{B_5}{B_4}, & \text{inside, }(r<R) \\ 
1, & \text{outside, }(r>R)
\end{array}
\right\}  \label{b1}
\end{equation}

\noindent We therefore have $\nabla \cdot \vec{D}=j^0=\rho $, where $\vec{D}%
=\kappa \vec{E}$, and by Gauss' Law the radial displacement field is $(4\pi
r^2)D=q_{en}=\int_V\rho dV$, and $q_{en}=Q_{en}e$, where $Q_{en}$ is the
number of bosons enclosed within the volume $V$. The radial component of the
displacement field is therefore 
\begin{equation}
D=\kappa E(r)=\frac{q_{en}(r)}{4\pi r^2}=\frac{Q_{en}(r)e}{4\pi r^2}
\label{e23}
\end{equation}

\noindent Inside the bubble we approximate the charge density as a constant
so that $\rho =Qe/(\frac 43\pi R^3)$, and 
\begin{equation}
q_{en}(r)=\left\{ 
\begin{array}{cc}
%TCIMACRO{\dfrac{Qer^3}{R^3} }
%BeginExpansion
{\displaystyle {Qer^3 \over R^3}}
%EndExpansion
, & (r\leq R) \\ 
Qe, & (r\geq R)
\end{array}
\right\}  \label{e23a}
\end{equation}

\noindent We therefore obtain the displacement field

\begin{equation}
D=\left\{ 
\begin{array}{ll}
\kappa _5E_5=%
%TCIMACRO{\dfrac{Qer}{4\pi R^3} }
%BeginExpansion
{\displaystyle {Qer \over 4\pi R^3}}
%EndExpansion
, & (r<R) \\ 
E_4=%
%TCIMACRO{\dfrac{Qe}{4\pi r^2} }
%BeginExpansion
{\displaystyle {Qe \over 4\pi r^2}}
%EndExpansion
, & (r>R)
\end{array}
\right\}  \label{e24}
\end{equation}

\noindent where $E_5$ ($E_4$) denotes the electric field inside (outside)
the bubble. The displacement $\vec{D}$ is continuous at the bubble wall, but
with our thin wall approximation with an infinitely thin domain wall, the
value of the electric field jumps up by a factor of $\kappa _5$ on the outer
surface of the bubble. The interior of the bubble appears as a dielectric
with an enormous dielectric constant $\kappa _5\gg 1$.

\subsection{Electromagnetic Energy ${\cal E}_{em}$}

We calculate the EM energy associated with the configuration by integrating
the EM energy density $u=\frac 12\vec{D}\cdot \vec{E}$ over all space. From (%
\ref{e24}) 
\begin{equation}
u=\left\{ 
\begin{array}{cc}
\left( 
%TCIMACRO{\dfrac{Q^2\alpha }{8\pi } }
%BeginExpansion
{\displaystyle {Q^2\alpha  \over 8\pi }}
%EndExpansion
\right) 
%TCIMACRO{\dfrac{r^2}{\kappa _5R^6} }
%BeginExpansion
{\displaystyle {r^2 \over \kappa _5R^6}}
%EndExpansion
, & (r<R) \\ 
\left( 
%TCIMACRO{\dfrac{Q^2\alpha }{8\pi } }
%BeginExpansion
{\displaystyle {Q^2\alpha  \over 8\pi }}
%EndExpansion
\right) 
%TCIMACRO{\dfrac 1{r^4} }
%BeginExpansion
{\displaystyle {1 \over r^4}}
%EndExpansion
, & (r>R)
\end{array}
\right\}  \label{e30}
\end{equation}

\noindent where $\alpha =e^2/4\pi $. The EM configuration energy 
\begin{equation}
{\cal E}_{em}=\int_0^\infty u\,4\pi r^2dr=2\pi \int_0^R\kappa
_5E_5^2\,r^2dr+2\pi \int_R^\infty E_4^2\,r^2dr  \label{e30a}
\end{equation}
\noindent is then given by

\begin{equation}
{\cal E}_{em}=\frac{Q^2\alpha }{2R}\left( 1+\frac 1{5\kappa _5}\right) =%
\frac{Q^2\alpha }{2R}\left( 1+\frac 15\frac{B_4}{B_5}\right)  \label{e31}
\end{equation}

\noindent For $\kappa _5\gg 1$, the contribution to ${\cal E}_{em}$ from the
interior region is negligible in comparison to the contribution from the
exterior region. If the dielectric constant inside the bubble were unity,
the EM energy would be $\frac 35\frac{Q^2\alpha }R$, so that the effect of a
macroscopic extra dimension is a reduction of the EM energy of the bubble.

\subsection{Bubble Mass and Radius}

Eq. (\ref{e22}) for the bubble mass gives, approximately, 
\begin{equation}
M={\cal E}_\chi +{\cal E}_{em}+{\cal E}_W+{\cal E}_V=\frac{Q\pi }R\left( 1+%
\frac{Q\alpha c}{2\pi }\right) +4\pi \sigma R^2+\frac 43\pi \lambda R^3
\label{e34}
\end{equation}

\noindent where 
\begin{equation}
c=\left( 1+\frac 1{5\kappa _5}\right) =\left( 1+\frac 15\frac{B_4}{B_5}%
\right)  \label{e34a}
\end{equation}

\noindent The equilibrium radius is obtained by minimizing the expression
for $M$ with respect to $R$, holding the charge $Q$ fixed. The equilibrium
mass of a bubble at its equilibrium radius can then be obtained. We consider
two limiting cases allowing us to obtain analytical expressions for the
bubble's equilibrium mass and radius: (i) the surface term ${\cal E}_W$ is
negligible and (ii) the volume term ${\cal E}_V$ is negligible.

{\it (i) Negligible Surface Term }${\cal E}_W$: For the first case we assume
that ${\cal E}_W\ll {\cal E}_\chi +{\cal E}_{em}$ and ${\cal E}_W\ll {\cal E}%
_V$, which can be rewritten as the conditions 
\begin{equation}
\frac{4\sigma R^3}{Q\left( 1+\frac{Q\alpha c}{2\pi }\right) }\ll
1,\,\,\,\,\,\,\,\,\,\,\frac{3\sigma }{\lambda R}\ll 1  \label{e34b}
\end{equation}

\noindent In this case the bubble mass is approximately $M={\cal E}_\chi +%
{\cal E}_{em}+{\cal E}_V$. We get for the equilibrium radius $R$ and
equilibrium mass $M$, respectively, 
\begin{equation}
R=\left[ \frac{Q\left( 1+\frac{Q\alpha c}{2\pi }\right) }{4\lambda }\right]
^{1/4}  \label{e34c}
\end{equation}

\noindent and 
\begin{equation}
M=\pi (4\lambda )^{1/4}\left[ Q\left( 1+\frac{Q\alpha c}{2\pi }\right)
\right] ^{3/4}=\frac{\pi Q}R\left( 1+\frac{Q\alpha c}{2\pi }\right) 
\label{e34d}
\end{equation}

\noindent Using (\ref{e34c}) we find that the conditions of (\ref{e34b}) are
approximately satisfied for 
\begin{equation}
\sigma \ll \left[ \lambda ^3Q\left( 1+\frac{Q\alpha c}{2\pi }\right) \right]
^{1/4}  \label{e34e}
\end{equation}

{\it (ii) Negligible Volume Term }${\cal E}_V${\it :} For the second case we
assume that ${\cal E}_V\ll {\cal E}_\chi +{\cal E}_{em}$ and ${\cal E}_V\ll 
{\cal E}_W$, which can be rewritten as the conditions 
\begin{equation}
\lambda \ll \frac{3\sigma }R,\,\,\,\,\,\,\,\,\,\,\lambda \ll \frac{3Q}{4R^4}%
\left( 1+\frac{Q\alpha c}{2\pi }\right)   \label{e34bb}
\end{equation}

\noindent The bubble mass in this case is approximately $M={\cal E}_\chi +%
{\cal E}_{em}+{\cal E}_V$, and we get for the equilibrium radius $R$ and
equilibrium mass $M$, respectively, 
\begin{equation}
R=\frac 12\left[ \frac Q\sigma \left( 1+\frac{Q\alpha c}{2\pi }\right)
\right] ^{1/3}  \label{e34cc}
\end{equation}

\noindent and 
\begin{equation}
M=3\pi \sigma ^{1/3}\left[ Q\left( 1+\frac{Q\alpha c}{2\pi }\right) \right]
^{2/3}=\frac{3\pi Q}{2R}\left( 1+\frac{Q\alpha c}{2\pi }\right)
\label{e34dd}
\end{equation}

\noindent Using (\ref{e34cc}) we find that the conditions of (\ref{e34bb})
are approximately satisfied for 
\begin{equation}
\lambda \ll \sigma ^{3/4}\left[ Q\left( 1+\frac{Q\alpha c}{2\pi }\right)
\right] ^{-1/3}  \label{e34ee}
\end{equation}

We can notice that the ratio 
\begin{equation}
\frac{{\cal E}_{em}}{{\cal E}_\chi }=\frac{Q\alpha c}{2\pi }=\frac{Q\alpha }{%
2\pi }\left( 1+\frac 1{5\kappa _5}\right)  \label{e37}
\end{equation}

\noindent indicates that ${\cal E}_\chi $ dominates ${\cal E}_{em}\,$ for $%
Q\ll 2\pi /\alpha $ and vice versa for $Q\gg 2\pi /\alpha $.

\section{Photon Opacity and Photon Stabilized Bubbles}

\subsection{Bubble Wall Opacity}

An interesting effect associated with the difference in space
dimensionalities inside and outside of a dimension bubble is the opacity of
the bubble wall to electromagnetic radiation. From the Lagrangian for the EM
field $F_{\mu \nu }$, given by (\ref{a4}), we have the field equations $%
\nabla _\mu (\kappa F^{\mu \nu })=j^\nu $, which represents Maxwell's
equations in terms of the displacement field $\vec{D}=\epsilon \vec{E}%
=\kappa \vec{E}$ and the magnetic field $\vec{H}=\vec{B}/\mu =\kappa \vec{B}$%
. We then identify the permittivity $\epsilon $, permeability $\mu $, index
of refraction $n=\sqrt{\epsilon \mu }$, and ``impedance'' $Z=\sqrt{\mu
/\epsilon }$ of a region of space as 
\begin{equation}
\begin{array}{ccc}
\epsilon =%
%TCIMACRO{\dfrac 1\mu }
%BeginExpansion
{\displaystyle {1 \over \mu}}
%EndExpansion
=\kappa ,\,\,\,\,\, & n=\sqrt{\epsilon \,\mu }=1,\,\,\,\,\, & Z=\sqrt{\mu
/\epsilon }=%
%TCIMACRO{\dfrac 1\kappa}
%BeginExpansion
{\displaystyle {1 \over \kappa}}
%EndExpansion
\end{array}
\label{c1}
\end{equation}

\noindent The reflectivity and transmissivity of light at an interface
between two dielectrics with indices $\epsilon _I$, $\mu _I$ and $\epsilon
_T $, $\mu _T$, where $I,T$ represent the incident and transmitting media,
respectively, can be obtained from classical electrodynamics. For example,
light impinging upon an interface between two media with impedances $Z_I$
and $Z_T$ has an associated reflection ratio given\footnote{%
This result for the reflection ratio ${\cal R}$ is true for all angles of
incidence. This follows from Snell's law and the fact that the index of
refraction is unity everywhere.} by\cite{em} 
\begin{equation}
{\cal R}=\frac{E_R}{E_I}=\frac{1-(Z_T/Z_I)}{1+(Z_T/Z_I)}  \label{c2}
\end{equation}

\noindent For light impinging upon the bubble wall from either the inside or
the outside, we have $|{\cal R}|\approx 1-O(\kappa _5^{-1})\approx 1$, where 
$\kappa _5^{-1}=B_4/B_5\ll 1$. Therefore the bubble wall is almost totally
opaque to EM radiation, and photons inside the bubble remain effectively
trapped inside the bubble. Photon radiation pressure therefore serves as yet
another means of stabilizing a dimension bubble against collapse.

\subsection{Photon Stabilized Bubble}

A dimension bubble may contain a bath of photon radiation in addition to the
charged bosons in its interior. Let us focus on a limiting case in which
essentially all the energy density of the bubble's contents is due to
photons. The photon energy density\footnote{%
In keeping with our Kaluza-Klein zero mode ansatz, we assume that there are
essentially no nonzero momentum states in the direction of the extra
dimension.} is $\rho _\gamma =(\pi ^2/15)T^4=aT^4$, so that the energy of
the bubble is 
\begin{equation}
M={\cal E}_\gamma +{\cal E}_W=\frac 43\pi \rho _\gamma R^3+4\pi \sigma R^2=%
\frac{4\pi ^3}{45}T^4R^3+4\pi \sigma R^2  \label{c3}
\end{equation}

\noindent The bubble, after it forms, will adjust its radius to reach an
equilibrium with an adiabatic (isentropic) expansion or contraction. The
photon entropy density is $s_\gamma \sim T^3$, and if we assume that the
bubble adjusts its radius to reach equilibrium on a very small time scale so
that essentially no energy is lost from the bubble during equilibration, we
have that 
\begin{equation}
RT=const  \label{c4}
\end{equation}

\noindent Using these expressions, a minimization of the mass function gives
an equilibrium radius $R$ and an equilibrium mass $M$ of 
\begin{equation}
R=\frac{90}{\pi ^2}\frac \sigma {T^4},\,\,\,\,\,\,\,\,\,\,M=12\pi \sigma R^2
\label{c5}
\end{equation}

\noindent (One could also obtain these results by balancing the photon
pressure with the pressure associated with the tension in the domain wall.)

To estimate the lifetime of such a bubble, we use the fact that photons
slowly leak out through the bubble wall at a rate that depends upon the
transmission coefficient ${\cal T}\sim O(\kappa _5^{-1})$. Using Poynting's
theorem relating the Poynting flux through the bubble wall to the rate of
energy decrease inside the bubble, we estimate that the bubble loses energy
at a rate 
\begin{equation}
\frac{dM}{dt}\approx -S_T\,(4\pi R^2)  \label{c6}
\end{equation}

\noindent where $S_T={\cal T}S_{inc}={\cal T}\rho _\gamma $ is the Poynting
flux transmitted through the bubble wall, ${\cal T}$ is the transmission
coefficient, $S_{inc}$ is the Poynting flux incident upon the wall, and $%
\rho _\gamma =aT^4=(\pi ^2/15)T^4$ is the photon energy density. We then
have, from (\ref{c5}) and (\ref{c6}), 
\begin{equation}
\frac{dM}{dt}\approx -\left( {\cal T}aT^4\right) (4\pi R^2)\approx 24\pi
\sigma R\frac{dR}{dt}  \label{c7}
\end{equation}

\noindent which gives 
\begin{equation}
\frac{dR}{dt}\approx -{\cal T}\sim -O(\kappa _5^{-1})  \label{c8}
\end{equation}

\noindent At time $t$ a bubble has a radius $R\approx R_0-{\cal T}t$, and
the lifetime of the bubble is 
\begin{equation}
\tau \approx \frac{R_0}{{\cal T}}\sim O(\kappa _5)R_0  \label{c9}
\end{equation}

A remark is in order for the case of a dimension bubble with a 5d interior,
whose domain wall arises, in part, from the low temperature Rubin-Roth
potential contribution to the effective potential $U(\varphi )$. In this
case the local minimum of $U(\varphi )$ disappears at high temperatures, so
that the domain wall itself disappears, or ``bursts'', at some temperature $%
T_c$. Therefore the photon temperature of the bubble's (5d) interior must be
restricted to values $T\leq T_c$. At a temperature $T\sim T_c$ the bubble
would burst, releasing all of its radiation. From (\ref{c5}) we see that as
the bubble shrinks, the photon temperature increases. When the temperature
reaches $T\sim T_c$, the bubble explodes, so that the bubble lifetime is
actually $\tau \lesssim O(\kappa _5)R_0$.

\section{Summary}

The dimension bubble scenario has been extended to include Maxwell fields
and sources. Beginning with a 5d Maxwell theory with sources, the reduction
to four dimensions leads to an effective 4d dilaton-Maxwell system that, in
the dielectric approach, leads to an interpretation of the extra dimensional
scale factor $B(x)$ as a dielectric function $\kappa (x)=B(x)/B_4$ that
takes drastically different values $\kappa _5\gg 1$ in the interior of a
bubble and $\kappa =1$ in the exterior. One result of this is that the
electromagnetic energy associated with charged scalar bosons confined to the
bubble's interior is reduced from what would be the case for a 4d interior.
This is due to the fact that the electromagnetic energy density $u=\frac 12%
\kappa E^2=D^2/2\kappa $ is greatly suppressed in a 5d interior with
dielectric constant $\kappa _5$.

Another, somewhat striking, result is that the bubble wall possesses a near
total reflectivity for $\kappa _5\gg 1$, making it opaque to photons
incident upon it from either side. Photons that are trapped within the
bubble at its time of formation cannot easily escape, so that a dimension
bubble can be held in a metastable state, supported by entrapped photons
alone. The photon temperature, however, must remain below the critical
temperature $T_c$ above which the domain wall disappears and an existing
bubble would burst. A bubble filled with photons slowly decreases in size,
with a resulting lifetime $\tau \lesssim O(\kappa _5)R_0$. As the bubble
shrinks, the temperature $T$ inside increases until it reaches a critical
temperature $T_c$, at which time the bubble explodes, releasing its
radiation contents.

\smallskip 

\

\end{document}